\documentclass[]{aa}  
\usepackage{graphicx}
\usepackage{txfonts}

\newcommand{\Teff}{\hbox{$T\sb{\rm eff}$}}          
\newcommand{\logg}{\hbox{$\log g$}}
\newcommand{\Msun}{\hbox{M$\sb{\odot}$}}

\begin{document}

   \title{Atmospheric parameters and carbon abundance for hot DB white
   dwarfs} 

   \author{D. Koester \inst{1} \and J. Provencal\inst{2} \and
     B.T. G\"ansicke\inst{3}} 

   \institute{Institut f\"ur Theoretische
             Physik und Astrophysik, Universit\"at Kiel, 24098 Kiel,
             Germany\\ 
     \email{koester@astrophysik.uni-kiel.de} 
     \and Department of Physics and Astronomy, 
          University of Delaware, Newark DE 19716, USA
     \and Department of Physics, University of Warwick, 
           Coventry CV4 7AL, UK}

   \date{Version 3, July 17, 2014}

\titlerunning{Hot DB white dwarfs}

\abstract {Atmospheric parameters for hot DB (helium atmosphere) white
  dwarfs near effective temperatures of 25\,000\,K are extremely
  difficult to determine from optical spectroscopy. The neutral He
  lines reach a maximum in this range and change very little with
  effective temperature and surface gravity. Moreover, an often
  unknown amount of hydrogen contamination can change the resulting
  parameters significantly. This is particularly unfortunate because
  this is the range of variable DBV or V777 Her stars. Accurate
  atmospheric parameters are needed to help or confirm the
  asteroseismic analysis of these objects.  Another important aspect
  is the new class of white dwarfs -- the hot DQ --  whose spectra are
  dominated by carbon lines. The analysis 
  shows that their atmospheres are pure carbon.  The origin
  of these stars is not yet understood, but they may have an
  evolutionary link with the hotter DBs, as studied here.}
{Our aim is to determine accurate atmospheric parameters and element
  abundances and study the implications for the evolution of
white dwarfs of spectral classes DB and hot DQ.}
{High-resolution UV spectra of five DBs were studied with model
  atmospheres.  We determined stellar parameters and abundances or
  upper limits of C and Si. These objects were compared with cooler DBs
  below 20\,000\,K.} 
{We find photospheric C and no other heavy elements -- with extremely
  high limits on the C/Si ratio -- in two of the five hot DBs. We
  compare various explanations for this unusual composition that
  have been proposed in the literature: accretion of interstellar or
  circumstellar matter, radiative levitation, carbon dredge-up from
  the deeper interior below the helium layer, and a residual stellar
  wind. None of these explanations is completely satisfactory, and the
  problem of the origin of the hot DQ remains an open question.}  {}

\maketitle

\section{Introduction}
DB white dwarfs have helium-dominated atmospheres, with occasional
minor pollution by hydrogen (DBA), heavy metals (DBZ), or carbon
(DQ). Recent large-scale studies of optical spectra were reported
by \cite{Voss.Koester.ea07} and \cite{Bergeron.Wesemael.ea11}. The
latter 
work gives an excellent account of the past two decades of literature
as well as a summary of open questions: the DB gap, the amount and
origin of the hydrogen contamination, the mass distribution, and
average masses of DBs versus those of the DAs (hydrogen-dominated
atmospheres). 

In our current study, the emphasis is on accurately determining
atmospheric parameters, effective temperatures, and surface
gravities. The fitting of optical DB spectra near 25\,000\,K with
theoretical models is extremely difficult. With mixing-length
parameters ML2/1.25 \citep[see, e.g.,][for the
  notation]{Fontaine.Villeneuve.ea81, Tassoul.Fontaine.ea90} -- the
preferred choice by the Montreal group -- the maximum strength of the
neutral helium lines is reached in this range, and a broad plateau is
formed, where the overall line strength is practically constant, with
only very subtle changes in the profiles. Additionally, there is the
problem of a possible hydrogen contamination if no hydrogen lines are
visible \cite[see, e.g.,][for a detailed
  discussion]{Bergeron.Wesemael.ea11}. This temperature region is
important because the variable DB or V777 Her stars are found around
25\,000\,K. Accurate parameters are needed to locate the instability
strip for this class and to determine whether it is ``pure'', that is, whether all
stars in this strip are pulsators. Spectroscopic temperatures and
masses are important for comparisons with asteroseismic results, or to help
chose the best asteroseismic models if many more pulsation frequencies
are predicted than observed.

Another important aspect for the hot DBs is the new class of white
dwarfs detected by \cite{Dufour.Liebert.ea07}, whose spectra are dominated
by \ion{C}{i} and \ion{C}{ii} lines. At first sight, they resembled
G\,35-26 and G\,227-5 at the hot end of the DQ spectral class around
12\,500\,K \citep{Thejll.Shipman.ea90, Wegner.Koester85}, but a
detailed first analysis showed that their atmospheres are dominated by
carbon instead of helium \citep{Dufour.Fontaine.ea08}. These are the
first objects in the ``normal'' white dwarf range below \Teff\ =
100\,000\,K, where the main element in the atmosphere is not hydrogen
or helium. Among the pre-white dwarfs at higher temperatures, the
PG\,1159 star H\,1504+65 is the only object that does not show any trace of H
or He in its spectrum \citep{Werner.Rauch.ea04}.

The hot DQs appear on the cooling track of white dwarfs around
24\,000\,K, which makes it very plausible that they are transformed from
another type, most likely the DBs \citep{Bergeron.Wesemael.ea11}.  It
is generally accepted that the cool DQ white dwarfs are formed by
carbon dredge-up, when the helium convection zone increases in depth
into the transition zone between helium and the underlying carbon
\citep{Koester.Weidemann.ea82, Pelletier.Fontaine.ea86,
  Dufour.Bergeron.ea05, Koester.Knist06}. It is therefore tempting to
assume a similar process for the hot DQs. The progenitors could be
stars like H\,1504+65, if they have retained a tiny amount of helium
in their outer layers -- a fraction 10$^{-15}$ of the total stellar
mass would be sufficient. In the subsequent cooling evolution this small
amount of helium would float to the top, transforming the star into a
helium-atmosphere white dwarf (DB) somewhere between
40\,000--30\,000\,K.  At 30\,000\,K a convection zone develops, which
deepens with decreasing effective temperature. If the He layer mass is
in the range of $10^{-15}$--$10^{-8}$ of the total mass, the
convection zone would reach into the He/C transition zone and rapidly
transform the star into a hot DQ. One aim of our study is therefore to
search for signs of such a transition in the hot DBs.

\section{Observations}
Our sample of five pulsating helium-atmosphere white dwarfs was chosen
with the goal of improving $T_\mathrm{eff}$ determinations. Our
targets were observed with the {\it Cosmic Origins Spectrograph} (COS)
onboard the {\it Hubble Space Telescope} (HST)
\citep{Green.Froning.ea12, Osterman.Green.ea11}.  Spectra were
obtained in TIME-TAG mode, using the primary science aperture and the
G130M and G160M gratings, with central wavelengths of 1291 and
1611\,\AA, respectively. Each exposure used multiple FP-POS positions
to improve the limiting signal-to-noise ratio (S/N) and minimize fixed-pattern noise. The
total exposure times were chosen to reach the desired S/N of 20, except for the faint object WD\,J1929+4447. In this
case, we limited the target S/N to 10. The effective wavelength range
of 1132-1789\,\AA\ ensured that the spectral coverage included Lyman
$\alpha$, \ion{O}{i}\,1302\,\AA, \ion{C}{ii}\,1335\,\AA, and
\ion{He}{ii}\,1640\,\AA. The wavelength resolution of
$\approx$0.1\,\AA\ allows the clear separation of interstellar and
photospheric line components.  Each data set was reduced with the COS
calibration pipeline, which includes corrections for flats, deadtime,
geometric and thermal distortion, and orbital and heliocentric Doppler
effects. The journal of observations is given in Table~\ref{observ}.

\begin{table*}[ht]
\caption{COS/HST log of observations for the five pulsating DB stars in the
  sample. The last column (opt) gives the optical photometry used for
  the parameter determination (Johnson $V$ or SDSS $g$).
\label{observ}} 
\begin{center}
\begin{tabular}{lrrrrrrr}
\hline
\noalign{\smallskip}
 target &     date & exposure [s] & aperture & grating & central
 $\lambda$ & opt \\ 
\noalign{\smallskip}
\hline
EC\,20058-5234 & 2011-11-19& 1100.64& PSA&  G130M& 1283.307& 15.58 (V) \\
               & 2011-11-19& 3423.58& PSA&  G160M& 1604.309 \\
WD\,J1929+4447   & 2012-11-09& 5182.72& PSA&  G130M& 1283.080& 18.21 (V) \\
                 & 2012-11-09& 5882.67& PSA&  G160M& 1604.282 \\
EC\,04207-4748   & 2012-10-24& 1419.14& PSA&  G130M& 1283.264& 15.30 (V) \\
                 & 2012-10-24& 2193.25& PSA&  G160M& 1604.418 \\
PG\,1115+158     & 2013-02-01& 1734.05& PSA&  G130M& 1283.322& 16.61 (g) \\
                 & 2013-02-01& 5246.78& PSA&  G160M& 1604.271 \\
WD\,1654+160     & 2013-07-04& 1409.08& PSA&  G130M& 1283.307& 16.62 (g) \\
                 & 2013-07-04& 2403.42& PSA&  G160M& 1609.345\\
\hline
\end{tabular}
\end{center}
\end{table*}

\begin{figure}
\centering
\includegraphics[width=0.45\textwidth]{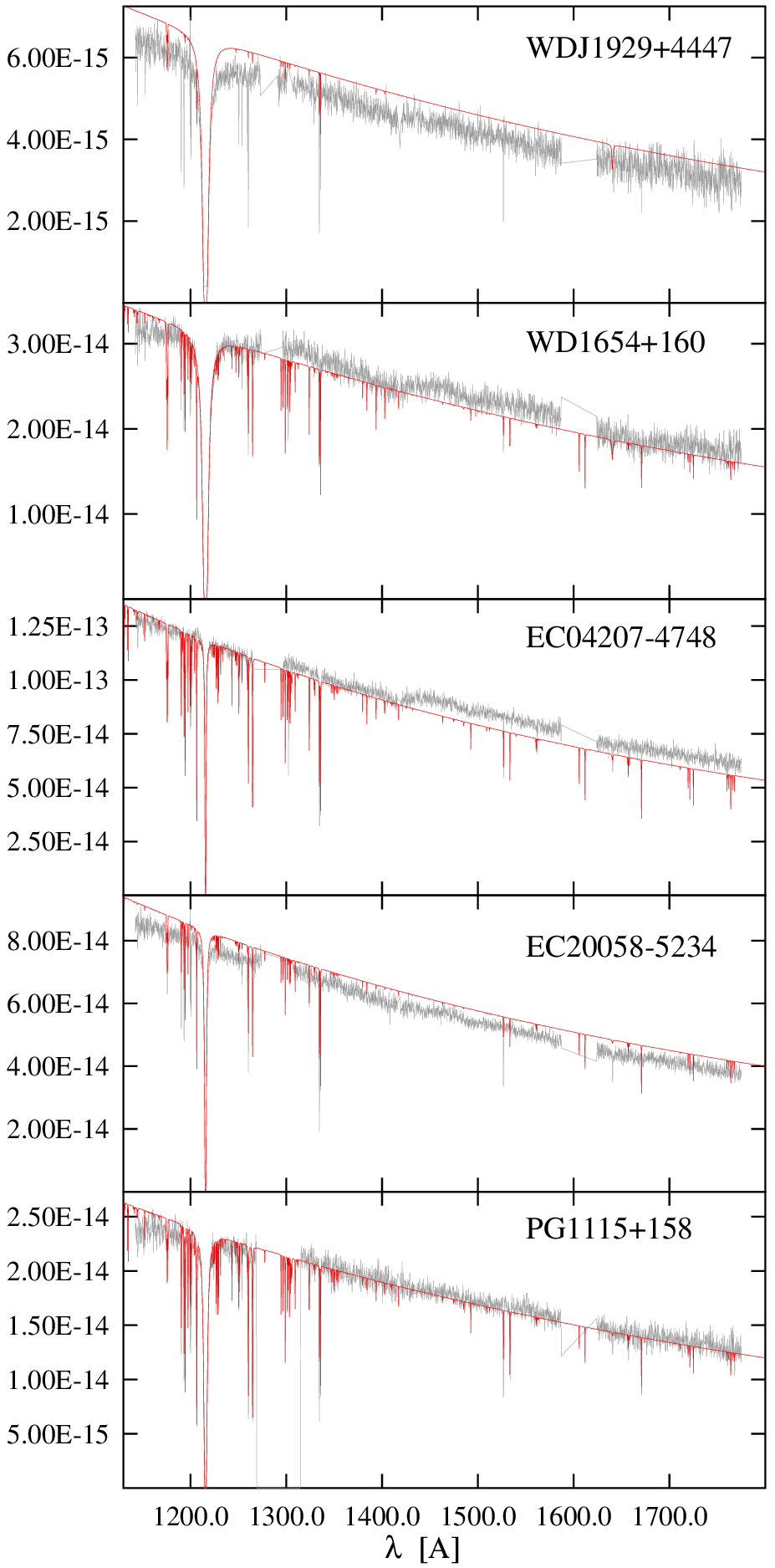}
\caption{HST/COS spectra of the five variable DBs (light gray),
  compared with model spectra (dark red). The vertical axis is the
  observed energy flux $F_{\lambda}$
  [erg\,cm$^{-2}$\,s$^{-1}$\,\AA$^{-1}$]. Note: the models are not fit
  to the observations, but scaled with the solid angle determined from
  the optical photometry. See text for more
  explanations. \label{spectra}}
\end{figure}

\begin{figure}
\centering
\includegraphics[width=0.45\textwidth]{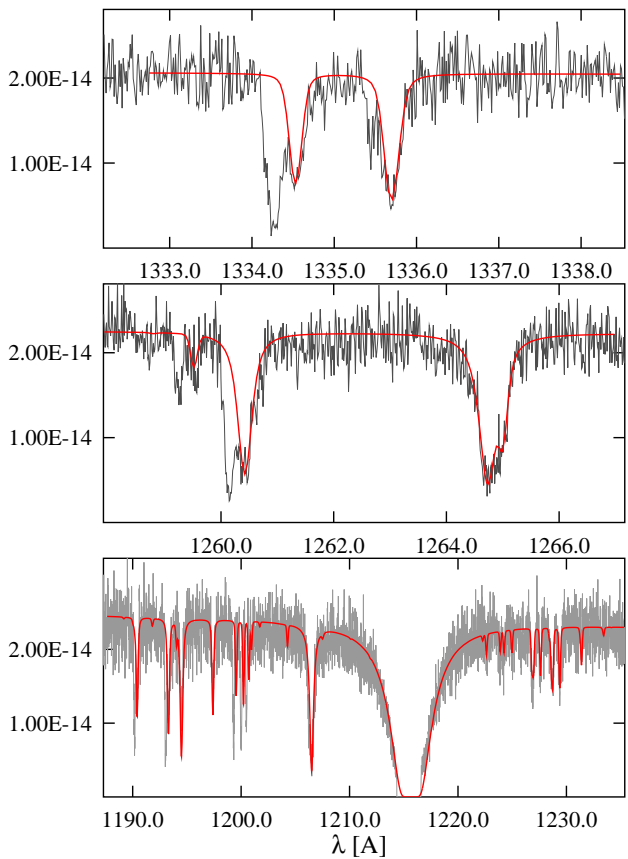}
\caption{Details in the PG\,1115+158 spectrum. Top panel:
  \ion{C}{ii}\,1330/1334\,\AA\ photospheric lines and blueshifted
  interstellar components. The vertical axis in all panels is the observed
  energy flux $F_{\lambda}$ [erg\,cm$^{-2}$\,s$^{-1}$\,\AA$^{-1}$]
  with zero at the bottom of each panel. Middle panel:
  \ion{Si}{ii}\,1260/1264/1265\,\AA\ photospheric lines and ISM
  component to the 1260\,\AA\ line. Bottom panel: Lyman $\alpha$,
  \ion{Si}{ii} (1190$-$1197, 1222$-$1229) and
  \ion{Si}{iii}\,1206\,\AA\ lines. The lines near 1199\,\AA\ are
  interstellar \ion{N}{i}. The continuous red line is the theoretical
  model.\label{PGdetails}}

\end{figure}

In addition, we used available optical spectra to redetermine the atmospheric
parameters. Spectra from the Supernova Progenitor Survey (SPY,
\cite{Voss.Koester.ea07}) were available for EC\,04207-4748, WD\,1654+160, and
PG\,1115+158. WD\,1654+160, and PG\,1115+158 were also recently analyzed by
\cite{Bergeron.Wesemael.ea11}. For WD\,J1929+4447 we used the excellent blue
optical spectrum analyzed in \cite{Oestensen.Bloemen.ea11}, which
does
not cover H$\alpha$, however. The same is true for the high S/N Magellan 6.5m telescope
spectrum of EC\,20058-5234 discussed in \cite{Sullivan.Metcalfe.ea07}. 

As a complementary sample of cooler, non-variable DBs we used seven
objects from the recent snapshot survey of white dwarfs, which covered the
temperature range of $\approx$17\,000--27\,000\,K
\citep{Gansicke.Koester.ea12, Koester.Gansicke.ea14}. This survey had DA
white dwarfs as primary targets, but included a few DBs for comparison
\citep{Farihi.Gansicke.ea13}. The HST/COS spectra were similar, but
were observed only with the G130M grating centered on 1291\,\AA, and
therefore extended only to 1450\,\AA. All these objects had been
analyzed using optical spectra by \cite{Bergeron.Wesemael.ea11}. We
used their parameters for our theoretical model and determined
metal abundances in the photosphere.

\section{Analysis of individual objects} 
PG\,1115+158 has a measured determination of
$\log(N(\mathrm{H})/N(\mathrm{He})) = -3.84$ (henceforth abbreviated
as [H/He]) \citep{Bergeron.Wesemael.ea11}. The other objects with
H$\alpha$ covered typically have limits of $-4.0$, those with only
H$\beta$ covered in the optical spectrum of $-3.5$ (WD\,J1929+4447,
EC\,20058-5234).

With a limit of [H/He] $< -4$ there is already an influence on
the spectra from the blanketing effect of the hydrogen Balmer and
Lyman jumps and L$\alpha$. In the optical range this is almost invisible
to the eye, but leads to significant changes in the position of the
minimum $\chi^2$ solution. Moreover, the $\chi^2$ surface becomes very
complicated, with two or even more local minima. The solution found by
the fitting code then depends on the starting conditions (we always
used 23\,000 and 28\,000\,K, which bracket the region of maximum line
strengths, and sometimes additional values as starting
points) and very minor details of the fitting procedure. The minimum
$\chi^2$ values are often identical, and at an S/N
below $\approx$30 the fits cannot be distinguished.

It is therefore desirable to use additional information beyond the
optical spectra. Trigonometric parallaxes are not available, but we
have the absolutely calibrated HST/COS spectra in the range 1140 --
1770\,\AA. The slope of these spectra mostly depends on the effective
temperature, which is even more true for the slope from the visual
to the UV. To give an example, at \Teff\ = 25\,000\,K a change of
$\Delta \logg\ = 0.25$, or of the hydrogen content [H/He] from -4 to  
-5 change the flux at 1400\,\AA\ by 1\%, whereas a change of 1000\,K
changes the flux by 9\%.

We therefore used the following steps in our analysis:
\begin{itemize}
\item optical spectra (described below) are fitted with a grid
  of helium-rich models and [H/He] of -5, using a $\chi^2$
  minimizing technique. The best solution is used as a starting point
  for the following steps:
\item the best-fit model is compared with the observed UV spectrum and
  the interstellar hydrogen column density $N$(HI) is determined from
  L$\alpha$. This is easily possible because in all objects this
line is
  either much stronger than predicted by the photospheric
  determination or limit, or clearly on the interstellar and not
  photospheric wavelength scale. From this column density we determine
  the optical extinction using the relation \citep{Bohlin.Savage.ea78,
    Gudennavar.Bubbly.ea12}
\begin{eqnarray}
  E(B-V) &=& \frac{N(\mathrm{HI})}{4.8\,10^{21}} \\
  A_V &=& 3.1 E(B-V)
.\end{eqnarray}
All derived values are lower than the highest values in the dust
maps of \cite{Schlegel.Finkbeiner.ea98}.
The dust extinction in the UV is modeled simply as linear in
$1/\lambda$ 
\begin{equation}
       A(\lambda) = \frac{5500 A_V}{\lambda}
,\end{equation}
which is an excellent fit for small $E(B-V)$ \citep{Zagury13}.

\item theoretical colors ($B$, $V$, $g$) are calculated for the
  best-fit models and compared with available photometry, corrected
  for extinction. From this comparison follows the solid angle of the
  star.

\item the theoretical UV spectrum is multiplied with the solid angle
and compared with the observed spectrum, regarding the slope in the UV
region, and the absolute values (i.e., the slope UV - $V$).
If the fit is not satisfactory, \Teff\ is changed, a new \logg\ is
found by spectroscopic fitting with \Teff\ fixed, and the iteration is
continued with the second entry of this list.
\end{itemize}

We used the [H/He] = -5 grid and verified that results using a
grid with [H/He] = -4 do not differ by more than a small fraction of
the errors. This is a big advantage of our method because four of the five
objects have only upper limits on the order of [H/He] $\approx -4$.

After the parameters effective temperatures and surface gravities
 were fixed, the lines in the UV spectra were identified and
abundances determined. The results are presented in
Table~\ref{parameters}.

\begin{table*}[ht]
\caption{Atmospheric parameters \Teff, \logg, and C and Si abundances. For
  the hot variables we also give the interstellar hydrogen column
  density from L$\alpha$, which was used in deriving our
  parameters. Numbers in brackets are estimated errors. The distance
  is determined from the spectroscopic radius and the solid
  angle. The parameters for the cool objects are from
  \cite{Bergeron.Wesemael.ea11}, where the errors and more details
  can be found.}
\label{parameters}
\begin{center}
\begin{tabular}{lrrrrrrr}
\hline
\noalign{\smallskip}
 object      &      \Teff [K] & \logg &[C/He]& [Si/He]& $\log N$(HI)
 [cm$^{-2}$] & distance [pc]\\
\noalign{\smallskip}
\hline
WD\,J1929+4447 & 30000 (1000) & 7.89 (0.15)&$<-6.2$ &$ <-7.5$& 20.4 (0.1) & 395\\
WD\,1654+160   & 29410 (500)  & 7.97 (0.08)&$<-6.4$ &$ <-7.0$& 20.4 (0.1) & 170\\
EC\,04207-4748 & 25970 (500)  & 7.79 (0.08)& -5.6 (0.2)& $<-8.4$  & 18.9 (0.1) & 108\\
EC\,20058-5234 & 25500 (500)  & 8.01 (0.05)& -5.3 (0.2)& $<-8.7$  & 19.1 (0.1) & 104\\
PG\,1115+158   & 25000 (500)  & 7.91 (0.07)& -5.6 (0.2)& -6.3 (0.2)&19.8 (0.1) & 192\\
\hline
\noalign{\smallskip}
WD\,0100-068 &  19800 & 8.07 &$  -7.3$ & $  -7.5$\\
WD\,0435+410 &  16810 & 8.19 &$ <-9.0$ & $  -7.2$\\
WD\,0840+262 &  17770 & 8.30 &$ <-8.3$ & $ <-9.0$\\
WD\,1557+192 &  19570 & 8.15 &$ <-7.8$ & $ <-8.1$\\
WD\,1822+410 &  16230 & 8.01 &$  -7.2$ & $  -7.4$\\
WD\,1940+374 &  16630 & 8.07 &$ <-8.5$ & $ <-9.0$\\
WD\,2144-079 &  16340 & 8.18 &$ <-8.5$ & $  -7.8$\\
\hline
\end{tabular}
\tablefoot {PG\,1115+158 has additional photospheric S (-6.7$\pm$0.3) and Al
(-7.0$\pm$ 0.3). All hot objects show interstellar lines from H, C,
N, Si, S; some also O and/or Al.}
\end{center}
\end{table*}

\subsection{Notes on individual objects}
{\bf WD\,J1929+4447}: the best fit using only the blue optical
spectrum is 24\,950\,K, \logg = 8.029, similar to the results given in
\cite{Oestensen.Bloemen.ea11}. The reddening $E(B-V)$ derived from the
HI column density is 0.052, consistent with the value 0.05 used in
that paper in the fit for the optical spectrum. However, the USNO-A2.0
catalog \citep{Monet98} has the photographic magnitudes $B$ = 17.4 and
$R$ = 18.8. Using the relation from Eric Mamajek (2010)
\footnote{\url{http://www.pas.rochester.edu/~emamajek/memo_USNO-A2.0.html}}
\begin{equation}
       V = 0.429 + 0.325\, B + 0.645\, R
,\end{equation}
we derive $V$ = 18.21. The relation has significant scatter ($\sigma =
0.35$), the $V$ value is therefore quite uncertain. If we nevertheless
use it to derive a solid angle from the comparison with theoretical
models, we need a much higher effective temperature of 30\,000\,K to
reproduce the slope of the UV spectrum, which is independent of the
optical photometry. The absolute scale (UV-$V$) is $\approx$10\% too
high and would be best fitted around 28500\,K.  Because of the large
uncertainty in the $V$ magnitude, we prefer the higher temperature. The
HeII\,1640\,\AA\ line is also fitted well, and the optical fit is not
noticeably worse than at the lower \Teff.  We note
that the pulsational properties also argue for a position of this
object at the blue edge of the instability strip near 29\,200\,K
\citep{Bischoff-Kim.Oestensen11}.

Only lines of the interstellar medium (ISM) are visible in the
UV. The upper limit on hydrogen from H$\beta$ is [H/He] = -3.5 (the
spectrum does not extend to H$\alpha$), but even this abundance is far
too small to explain L$\alpha$. This line clearly is on the ISM
wavelength scale and dominated by interstellar absorption.

{\bf WD\,1654+160}: We have two SPY spectra of relatively low
quality. The optical fits result in low \Teff\ values
($\approx$23\,000\,K) with very high uncertainty. We prefer to use the
much higher result (see Table~\ref{parameters}) from
\cite{Bergeron.Wesemael.ea11} (with [H/He] $<$ -3.98), which
is based on much
better spectra. This model, with $V$ = 16.55 \citep{Farihi.Becklin.ea05}
and an extinction calculated from the hydrogen column density, predicts
the correct UV slope. The absolute value of the UV flux (or the UV-$V$
slope) is slightly too low, but compatible within the
errors. WD\,1654+160 also has photometry from the SDSS survey
\citep{Aihara.Allende-Prieto.ea11}; the value of $ g = 16.36$ (16.16
after dereddening) results in an excellent fit of the UV spectrum.  The
\ion{He}{ii}\,1640\,\AA\ line agrees with this model. In the UV spectrum
only ISM lines are visible. [H/He] = -4, the limit from the optical
spectra, is far too low for L$\alpha$, which must be dominated by the
ISM absorption.

{\bf EC\,04207-4748}: We used two SPY spectra of high S/N (40 and 30),
with a well-defined spectroscopic fit at \Teff\ = 25\,970, \logg\ =
7.790, and [H/He] = -5.0. The UV spectrum has an artificial jump
near 1400\,\AA, which makes the comparison difficult. With $V$ = 15.3
\citep{Kilkenny.ODonoghue.ea09}, the average UV-$V$ slope is best fitted
at 26\,500\,K, but then the slope below 1400\,\AA\ in the UV is too
steep. We keep the spectroscopic solution in this case.

The UV spectrum shows interstellar lines of several elements and
photospheric lines only of carbon. \ion{C}{ii}\,1323\,\AA\ is a highly
excited line and must be photospheric. On this wavelength scale
\ion{C}{ii}\,1335\,\AA\ and the \ion{C}{iii} lines near 1175\,\AA\ are
also excellent fits. There are obviously two blue ISM components to
\ion{C}{ii}\,1334\,\AA. These two components are also visible in the
\ion{Si}{ii}\,1260\,\AA\ and the \ion{S}{ii} lines. The
\ion{C}{iii}/\ion{C}{ii} ionization ratio is well reproduced at this
temperature. A hydrogen abundance of [H/He] = $-3.7$ would fit
L$\alpha$ and is consistent with an upper limit of $-3.5$ from the
optical H$\beta$. However, L$\alpha$ is shifted on the ISM wavelength
scale, and the photospheric abundance is certainly lower.

{\bf EC\,20058-5234 (QU Tel)}: We used a high-quality spectrum taken
at the Magellan 6.5m telescope and discussed in
\cite{Sullivan.Metcalfe.ea07}. With the new model grid we find the
best spectroscopic solution at 25\,000\,K, \logg\ = 8.01 and [H/He] =
-5.0, although local minima also exist at higher temperatures. Using
$V$ = 15.58 from \cite{Koen.ODonoghue.ea95}, the UV-$V$ slope is
reproduced correctly, but the UV slope is slightly too flat. It agrees
with a model at 26\,000\,K, but then the UV-$V$ slope is too steep. We
therefore use the compromise of 25\,500\,K and include both solutions
within the error. The \ion{He}{ii}\,1640\,\AA\ line is stronger than
predicted.

In addition to many interstellar lines, there clearly is photospheric carbon,
with lines from exited states of \ion{C}{ii} and \ion{C}{iii}.  The
\ion{C}{iii}/\ion{C}{ii} ratio points to a slightly lower
temperature. The upper limit from L$\alpha$ is [H/He] = -4, which would
fit the observations in strength, but is clearly shifted exactly as
the other ISM lines.

\cite{Petitclerc.Wesemael.ea05} studied {\it Far Ultraviolet Spectral
  Explorer} (FUSE) spectra of EC\,20058-5234. The only heavy
  element they found is carbon, their abundance (-5.50) agrees with our
  result within the errors. They also determined an upper limit for the
  Si/C ratio, which is very low, but a factor of two higher than our
  determination from COS spectra of superior quality.

{\bf PG\,1115+158 (DT Leo)}: In this case we also preferred the result
from \cite{Bergeron.Wesemael.ea11} (\Teff\ = 23\,770$\pm$1632\,K,
\logg\ = 7.91$\pm$0.07,[H/He] =-3.84). However, using the accurate
SDSS magnitude $g$ = 16.61 (0.02), the predicted flux and slope in the
UV are significantly too low. Since the effective temperature in this
case has a very large error, we have kept the surface gravity value
and iteratively increased \Teff. With 25\,000\,K the UV slope as well
as the UV-$V$ slope are accurately matched. The \ion{He}{ii
}\,1640\,\AA\ line is still predicted too weak, however.

In the UV, ISM lines of several elements are present and clearly
separated from photospheric lines, which include S and Al in addition
to C and Si.  The \ion{C}{iii}/\ion{C}{ii} and
\ion{Si}{iii}/\ion{Si}{ii} ratios confirm the temperature. The
hydrogen abundance from the optical determination is not nearly enough
to fit the L$\alpha$ line, which is also shifted to the ISM wavelength
scale and therefore must be dominated by interstellar absorption.

\subsection{General discussion of observations}
The adopted stellar parameters and abundances for C and Si are
summarized in Table~\ref{parameters}.  The UV spectra of the five
variable DBs are shown in Fig.~\ref{spectra} together with the
theoretical models for the parameters derived above. The model spectra
are corrected for interstellar extinction determined from the hydrogen
column density $N$(HI) and multiplied with the solid angle determined
from the optical photometry, compared with theoretical magnitudes for
the stellar parameters. We emphasize that the models were not
fitted to the UV observations, but represent the best compromise
between fits to optical spectra, UV spectral slope, and the UV-$V$
slope.  The models show more spectral lines than observed in the
stars, and also of elements other than C and Si (O, S, N, Fe). These
lines were included as wavelength references for the comparison with
interstellar lines, and to determine the photospheric nature of
carbon. A more detailed example for the photospheric and interstellar
lines of C and Si is shown in Fig.~\ref{PGdetails}.

We recall that all these stars are nonradial g-mode pulsators with
multiple modes of oscillations, which can experience temperature
excursions of 3000\,K during a pulsation cycle. Our UV spectra are
time-averaged spectra that cover multiple pulsation cycles. How well
our static models can fit the time-averaged observations depends
somewhat on periods, amplitudes, phases, and perhaps inclination. For
example, the large-amplitude pulsators EC\,04207-4748 and WD\,1654+160
present significantly nonlinear light curves with sharp maxima
generated by localized high-temperature variations
\citep{Montgomery.Provencal.ea10}. In particular the \ion{He}{ii}\,
1640\,\AA\ line could potentially show strong effects. A theoretical
nonpulsating model with \Teff\ = 25\,000\,K will show an extremely
weak (undetectable) line. However, the strength of the line rapidly
increases at higher temperatures, but is still 0 below
25\,000\,K. Thus a pulsating model can show a detectable line in the
time-averaged spectrum with time-averaged \Teff\ = 25000\,K, if the
amplitude is large enough. At the hot end near 29\,000\,K the change
is much more symmetric between positive and negative temperature
variations and the time-averaged spectrum will resemble the
nonpulsating star more closely. The observations were taken in
TIME-TAG mode, and we are currently working on extracting
time-resolved spectra to analyze the change during the pulsation
cycle.

\section{Origin of the photospheric metals}
The widely accepted explanation for the metal pollution in white dwarf
photospheres is accretion from a debris disk, a remnant of a former
planetary system, with rocky material resembling the composition of
our own system \citep[e.g.,][]{Zuckerman.Koester.ea07,
  Dufour.Kilic.ea12, Gansicke.Koester.ea12} . In a comprehensive study of
HST/COS spectra of a homogenous sample of DA white dwarfs with
\Teff\ between 17\,000 and 27\,000\,K, \cite{Koester.Gansicke.ea14}
found that 48 out of 85 objects showed photospheric metals, in all
cases including silicon. In about half of them, the presence of Si and
sometimes C could be explained by radiative levitation and no current
accretion is necessary. However, 23 objects must be currently
accreting because the abundances are too high to be supported by
radiative levitation. The diffusion timescales in these DAs are
extremely short, from days to a few years at most. It can therefore be
safely assumed that the accretion is now ongoing in a stationary
state, which allows the direct determination of the abundance ratios
in the accreted matter from those observed in the photosphere.

Only a few of the DAs with photospheric silicon also show carbon, and
the highest C/Si ratio for the accreting matter is 10 for
WD\,0710+471. This star is a known post-common-envelope binary
\citep[PCEB, see][]{Marsh.Duck96}, which accretes solar-like material
from the wind of its red companion. In all other cases the C/Si ratio
is at most around 1 and in most cases much below this, which is
typical for rocky material in our solar system. There is no single
object among the 48 with metals that shows carbon and not silicon.

A similar situation is seen in the seven cooler DBs below 20\,000\,K
that were included as comparison objects in the HST/COS study. Their
parameters are also given in Table~\ref{parameters}. Three of the
objects show no photospheric lines, two have a C/Si ratio of
$~\approx$ 1, and the other two have a ratio much lower than 1. Differently from the DAs, this sample is
strongly biased: the four DBs with metals were known before our study
to be polluted with metals, and were selected for that
reason. However, there was no bias in the selection concerning the
C/Si ratio. The origin of these metals is very likely accretion, as in
the DA sample.

\subsection{Accretion and diffusion in the hot DB}
A stunning difference is seen in the hotter DBs, with very high C
abundances and an observed photospheric C/Si number ratio $>630$ in
EC\,04207-4748 and $>2500$ in EC\,20058-5234 (the ratios by mass are
270 and 1076). These objects add to the growing number of hot DB stars
that show only carbon as heavy element, such as BPM\,17088, GD\,190
\citep{Petitclerc.Wesemael.ea05}, and PG0112+104
\citep{Dufour.Desharnais.ea10}, with similar carbon abundances and low
silicon limits. GD\,358 might be another example; it also shows
carbon. The GHRS ({\it Goddard High Resolution Spectrograph}) spectrum
shows separated ISM and photospheric \ion{C}{ii} lines
\citep{Dufour.Wesemael.ea02}, but does not cover the wavelength
regions of strong Si lines. The STIS ({\it Space Telescope Imaging
  Spectrograph}) spectrum we have \citep{Nitta.Koester.ea12}
shows carbon and also the Si 1260/1264\,\AA\ feature, but the
resolution is not high enough to clearly separate ISM and photospheric
contributions.

Differently from the DAs in this temperature range between
\Teff\ 20\,000--30\,000\,K, the DBs develop a convection zone. Since
the convection zones are always homogeneously mixed, diffusion starts
only from the bottom of this zone, leading to longer diffusion
times. Table~\ref{diffusion} summarizes the data on the convection
zones and diffusion timescales. Assuming that the metal pollution is
caused by accretion in a steady state with diffusion, we can derive
the diffusion fluxes or upper limits for C and Si, and thus the
abundances in the accreted matter. The derived C/Si mass ratios are
compared with the ratios in different objects of our own solar system,
showing the extraordinary discrepancy with those ratios in
EC\,04207-4748 and EC\,20058-5234.

The diffusion timescales range between 10 and 200 years. This is still
not very long, but one might argue that accretion is not in a
steady state, but has already stopped. Silicon, with shorter diffusion
timescales, could have already disappeared, while carbon is still
observable. A simple test calculation for EC\,04207-4748 shows that
this leads to rather extreme conclusions.  The abundance ratio C/Si
changes with time $t$ in years as
\begin{equation}
      \frac{N(\mathrm{C})}{N(\mathrm{Si})} = \left . 
       \frac{N(\mathrm{C})}{N(\mathrm{Si})} \right|_0 \exp(-t/179.67 +
        t/85.29) 
.\end{equation}
Assuming a starting ratio of 1, it would take 1271 years to reach the
observed lower limit. The starting abundance would have been [C/He] =
-2.23, with an accretion flow of $8.04\,10^9$ g/s of carbon
alone. Objects with such high accretion rates have never been observed,
and we consider this scenario as highly unlikely.

Before the accretion of planetary debris was commonly accepted as the
source of pollution in white dwarfs, accretion from comets was another
option discussed. The oxygen-to-carbon ratio in comets is close to the
solar value, [O/C] = 0.26. In EC\,04207-4748 the oxygen lines are
strongly perturbed by the geocoronal emission, but in EC\,20058-5234
we can set an upper limit [O/C]$<$-0.8, that is, about one tenth of the
solar value.

The maximum hydrogen content in the convection zone of
EC\,20058-5234 -- assuming [H/He]$<-3.5$ -- is
$2\times10^{17}$g. Scaling the carbon diffusion flux, using solar mass
ratios, the hydrogen accretion would be $5.8\times10^{16}$ g/yr; the
abundance change would be visible within a few months. These arguments
rule out comets as the source of the accretion.

\begin{table*}[ht]
\caption{Diffusion data: fractional mass in the convection zone,
  diffusion timescales $\tau$ for C and Si, and diffusion
  fluxes. $\Delta M(1)/M$ is the fractional mass above the layer with
  [C/He] = 1. In the last column the mass ratio C/Si of the accreted
  matter is given (with the assumption that the metal pollution is
  caused by accretion). The last three rows in this column give the
  mass ratio C/Si for the Sun \citep{Asplund.Grevesse.ea09}, bulk
  Earth \citep{Allegre.Manhes.ea01}, and CI chondrites
  \citep{Lodders.Palme.ea09}. \label{diffusion}}
\begin{center}
\begin{tabular}{lrrrrrr}
\hline
\noalign{\smallskip}
 object      & $\log(\Delta M_{cvz}/M) $ & $\log(\Delta M(1)/M)$&
 $\tau$(C) [yrs] & $\tau$(Si) [yrs]&  $f_{diff}$(C) [g/s] & C/Si \\
\noalign{\smallskip}
\hline
WDJ1929+4447 & -12.661 & $>$-10.962& 31.05 &  11.91 & $<4.67\,10^5$ \\
WD1654+160   & -12.466 & $>$-10.721& 36.23 &  15.14 & $<4.27\,10^5$ \\
EC20058-5234 & -11.656 &    -10.178& 114.89 &  55.77 & $ 5.90\,10^6$ &
                137.5\\
EC04207-4748 & -11.592 &    -10.045&179.67 &  85.29 & $ 6.81\,10^6$ &
                512.0\\
PG1115+158   & -11.373 & $>$-9.822& 204.64 & 103.40 & $ 5.52\,10^5$ &
                0.11\\
\hline
WD0100-068   &  -8.419 & $>$-6.062 &$1.82\,10^4$& $1.40\,10^4$&
                  $ 1.26\,10^6 $& 0.52\\
WD0435+410   &  -6.676 & $>$-3.627 &$2.43\,10^5$& $2.05\,10^5$&
                  $<9.36\,10^4$&$<0.005$ \\
WD0840+262   &  -7.263 & $>$-4.437 &$8.49\,10^4$& $6.81\,10^4$&
                  $<4.69\,10^5$&$<1.72$ \\
WD1557+192   &  -8.184 & $>$-5.681 &$2.36\,10^4$& $1.90\,10^4$&
                  $<5.63\,10^5$&$<0.68$\\
WD1822+410   &  -6.130 & $>$-3.644 &$7.42\,10^5$& $6.64\,10^5$&
                  $ 7.16\,10^6 $&$0.61$ \\
WD1940+374   &  -6.393 & $>$-3.499 &$4.57\,10^5$& $4.01\,10^5$&
                  $<3.35\,10^5 $& \\
WD2144-079   &  -6.482 & $>$-3.686 &$3.28\,10^5$& $2.81\,10^5$&
                  $<4.17\,10^5 $&$<0.07$ \\
\hline
Sun          &        &           &             &&  & 3.56   \\
bulk Earth   &        &           &             &&  & 0.0099 - 0.0228\\
CI           &        &           &             &&  & 0.325\\
\hline
\end{tabular}
\end{center}
\end{table*}

\subsection{Radiative levitation}
Radiative levitation for C and Si has recently been shown to be
important for DA white dwarfs around \Teff\ = 20\,000\,K
\citep{Chayer.Dupuis10, Dupuis.Chayer.ea10, Chayer14}. Similar
calculations do not yet exist for DB white dwarfs, but useful data can
be found in \cite{Chayer.Fontaine.ea95}, although the emphasis there
is more on the hotter white dwarfs. In their Fig.~16 the authors
present the equilibrium abundance in the envelope of DO/DB white
dwarfs (with \logg\ = 7.5) for many elements, including C and Si, for
temperatures from 100\,000 to 20\,000\,K. At a fractional depth between
$\Delta M/M = 10^{-13}$ and $10^{-8}$, significant abundances of many
elements can be supported by radiative levitation, at least at higher
temperature. Upon cooling below 30\,000\,K, the developing convection
zones could reach into these ``metal clouds'', mixing it homogenously
even up into the photosphere. For C, however, the abundances at
30\,000\,K are already off the scale of the figure, below [C/He] =
-10. At the higher surface gravities of our objects (>7.5), the
abundances would be even lower.  Equally important, many other
elements show similar behavior, and in particular silicon would be
supported at a significantly larger abundance than carbon. We
conclude that dredge-up from such radiatively supported metal
clouds cannot explain the observed C/Si ratios.

\subsection{Carbon dredge-up}
Another process that in principle can pollute a helium atmosphere
with carbon is dredge-up from the tail of the C-He transition region
formed after the end of helium-burning. This has been the standard and
well-understood explanation of the cool DQ stars below 12\,000\,K
\citep{Koester.Weidemann.ea82, Pelletier.Fontaine.ea86,
  Dufour.Bergeron.ea05, Koester.Knist06}. It was an obvious idea that
such a dredge-up could also explain the hot DQs, and this has been
speculated among others by \cite{Bergeron.Wesemael.ea11}. The problem
is that the helium layer in at least some DB stars needs to be much
thinner than the standard of about 1/100 of the total mass predicted
by stellar evolution calculations, and no indication for that has been
found so far. Nevertheless, in the absence of a generally
  accepted origin for the hot DQs, we investigate below necessary
  conditions for carbon dredge-up in the two objects, and the
  consequences.

\begin{figure}
\centering
\includegraphics[width=0.45\textwidth]{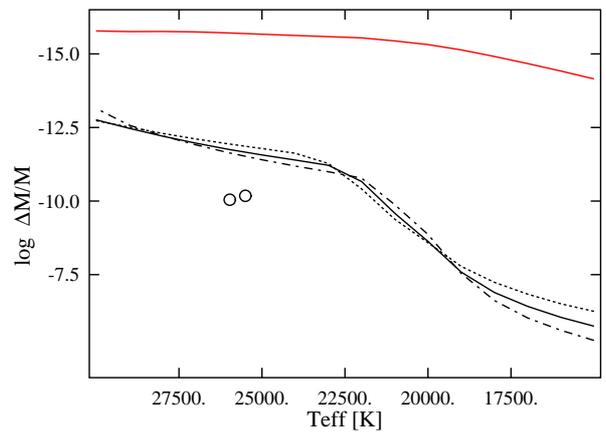}
\caption{Depth of convection zones with temperature. The upper
  continuous (red) line is the position of the photosphere at
  Rosseland $\tau =2/3$. The three lower curves are the bottom of the
  convection zone for \logg\ = 7.75, 8.00, 8.25. The two circles are
  the predicted depths with $N$(C)/$N$(He) = 1 for EC\,20058-5234 and
  EC\,04207-474.  \label{cvz}}
\end{figure}

\cite{Vennes.Pelletier.ea88} derived the equations (their Eqs.
11-22) governing the diffusion equilibrium between hydrogen and helium
in a general way, which can be directly applied in our case for
the helium-to-carbon transition as well. With $Z_1$ and $Z_2$ as average
charges of the two ions, and $A_1$ and $A_2$ the atomic weights, the
equations we need are
\begin{eqnarray}
\frac{\partial \gamma}{\partial \ln p} &=& \frac{E(\gamma)\,(A_2 Z_1 - A_1
  Z_2) + D(\gamma)\,(Z_2 -
  Z_1)}{F(\gamma)+G(\gamma)} \\
    F(\gamma) &=&  \frac{C(\gamma)\,D(\gamma)}{\gamma B(\gamma)} \\
    G(\gamma) &=&
    \frac{D(\gamma)\,(Z_2-Z_1)^2}{B(\gamma)\,E(\gamma)}\\
    B(\gamma) &=& 1+\gamma \\
    C(\gamma) &=& Z_1 + \gamma Z_2 \\
    D(\gamma) &=& A_1 + \gamma A_2 \\
    E(\gamma) &=& (1+Z_1) + \gamma (1+Z_2)
.\end{eqnarray}
Here $\gamma$ is the number ratio $n_2/n_1$, $p$ the pressure. The
nonlinear equation for $\gamma$ is integrated from the known starting
values at the bottom of the convection zone to a specified value,
for example, 1.0. The fractional mass at that depth $\Delta M/M$ can easily be
obtained from an integration of the hydrostatic equation assuming a
constant surface gravity $g$ -- a good approximation in the outer
layers
\begin{equation}
        q = \frac{\Delta M}{M} = \frac{4 \pi G}{g^2}\,p
.\end{equation}
The results of these calculations are also shown in
Table~\ref{diffusion}, where the third column gives the mass depth at
which the number ratio is 1.0, or with our logarithmic variable [He/C]
= 0. In all cases, where no photospheric carbon is observed, or were
accretion has occurred possibly in addition to dredge-up, these numbers
are lower limits, and the helium layer might have the canonical
thickness predicted by stellar evolution theory. However, for
EC\,20058-5234 and EC\,04207-4748 -- if the dredge-up scenario is 
correct -- a very thin helium layer is predicted. This will have
inevitable consequences for the further
evolution. Figure~\ref{cvz} shows the change of the depth of the
convection zone with decreasing effective temperature. At first more
gradually, but below 22\,500\,K rather rapidly, the bottom of this
zone reaches increasingly deeper layers. The two circles show the
predicted 
transition layer for the two objects with the extreme carbon
abundance; it is plausible that with further cooling near 22\,000\,K
they 
will become extremely carbon-rich, that is, hot DQs. 

While at first sight this appears to be a plausible explanation
  for the abundant carbon pollution and the hot DQs, there are severe
  problems that make accepting this hypothesis difficult.The
  most severe is the number statistics: finding two out of five DBs
  near 
  25\,000\,K in this study, as well as at least three more in the
  literature (see above) as possible hot DQ progenitors is clearly at
  odds with the fact that these stars seem to be extremely
  rare. \cite{Dufour.Liebert.ea07} estimated that only one out of 2850
  white dwarfs belongs to this class. The real comparison must
  be made with the number of DBs in the narrow range were the DQs
  exist, which is much lower, but a 40\% incidence seems impossible to
  explain. Another problem is presented by the masses. Although still
  preliminary, fits to spectroscopic observations suggest a high mass
  with surface gravities near 8.75 \citep{Dufour.Ben-Nessib.ea11};
  recent parallax measurements confirm this for two objects
  (private communication from the referee P. Dufour).

\subsection{Residual stellar wind}
Historically, the first hypothesis for the carbon pollution of DB
white dwarfs was proposed by \cite{Fontaine.Brassard05} and
\cite{Brassard.Fontaine.ea07} even before the hot DQs were
discovered. These authors assumed a PG1159-DO-DB-DQ evolutionary
sequence and a 
stellar wind that decreases from high effective temperatures linearly
with age and dies at 20000\,K. They stated themselves that this is an
ad hoc assumption without physical basis, and cannot be seen as a
continuation of the stellar winds at much higher \Teff\ discussed by
\cite{Unglaub.Bues00}. To our knowledge, there is no observational
evidence for such winds from white dwarfs. The standard assumption of
\cite{Fontaine.Brassard05} corresponds to a mass loss of
$1.3\times10^{-13}$\Msun/yr for a DB at 25000\,K, but calculations are
also performed with the wind increased by factors 2.5 and 5. This wind
would slow down the gravitational settling of carbon, thus explaining
the carbon in DBs above 20000\,K. Judging from their Fig.~1, the mass
loss to explain the carbon abundance in our two objects would have to
be $\approx 2\times10^{-13}$ \Msun/yr. This model did not attempt to
explain the hot DQs, which were not known at the time.

Apart from the unknown physical mechanism and missing observational
evidence, there are two other problems with this hypothesis. The mass
loss should be higher at higher temperatures, yet the two hottest
objects of our sample show no carbon, with upper limits significantly
lower than the abundances observed in the cooler objects. And
second, PG\,1115+158, with photospheric Si, S, and Al detected,
obviously is accreting. The proposed mass loss in the residual-wind
scenario amounts to $1.27\times10^{13}$ g/s, compared with a total
accretion flow scaled from the Si mass flow with bulk Earth abundances
of $3\times10^{7}$ g/s. It seems impossible that accretion could occur
under such circumstances. If accretion is not occurring now, it must
have occurred fairly recently because of the short diffusion
timescales. But then the mass loss was even stronger because it
decreases proportional to age in the \cite{Fontaine.Brassard05}
scenario.

\section{Discussion and conclusions}
We have presented improved stellar parameters for five hot DB stars,
using optical photometry and absolutely calibrated HST/COS spectra in
addition to optical spectra. Abundance patterns for Si and C in four
of the seven DBs below 20000\,K are typical for accretion from
circumstellar material, while no photospheric heavy elements are found
in the other three.  Of the hot DBs, only one shows photospheric lines
of C, Si, S, and Al, which are very likely due to accretion. The two
hottest 
objects do not show any photospheric lines. The remaining two objects,
EC\,20058-5234 and EC\,04207-4748, both near 26000\,K, show high
carbon abundances and an extremely high C/Si ratio, joining at least
three other DBs in the same temperature range with similar
properties. We discussed possible hypotheses about the origin of this
carbon and excluded accretion and radiative levitation. 

Carbon dredge-up is the only hypothesis that provides a single
explanation for the carbon abundances in the hot DBs and the formation
of the hot DQs. This seems more natural than assuming two unrelated
explanations. We acknowledge that this is not a new idea:
\cite{Bergeron.Wesemael.ea11} explicitely stated that the immediate
progenitors of the hot DQs must be among the hot DBs. It is clear that
this can only occur for thin helium layers. Dredge-up also provides a
natural explanation for the absence (or lower C abundances) in the two
hot objects near 30000\,K (the shallower He convection zone), although
a thicker helium layer would have the same effect.  It is compatible
with the appearance of other heavy elements in one of the three
objects at 25000\,K, which could be due to accretion. The most severe
argument against the connection with the hot DQs is currently the
number statistics: there are too many possible progenitors for the
rare hot DQs. Time-dependent calculations of the evolution of the
convection zone during the mixing process will be necessary to
determine whether all hot DBs with carbon traces really have to evolve
into hot DQs, or if this perhaps occurs only for the most massive
objects.

A residual stellar wind could explain the carbon pollution, but
requires an unrelated mechanism for the hot DQs.  It also faces
difficult problems: an unknown physical mechanism, no observational
confirmation, absence of carbon in some objects, even in the hottest
two of the sample, where the wind should be strongest, or apparent
accretion in spite of much larger mass loss. This means that
both
the carbon pollution of hot DBs and the origin of the hot DQs
unfortunately currently remain unexplained.

\begin{acknowledgements}
Based on observations made with the NASA/ESA Hubble Space Telescope,
obtained at the Space Telescope Science Institute, which is operated
by the Association of Universities for Research in Astronomy, Inc.,
under NASA contract NAS 5-26555. These observations are associated
with program \#12929 and \#12474.  The research leading to these
results has received funding from the European Research Council under
the European Union's Seventh Framework Programme (FP/2007-2013) / ERC
Grant Agreement n. 320964 (WDTracer).  D.K. thanks S.O.Kepler
of UFRGS (Porto Alegre, Brazil) for stimulating discussions and kind
hospitality during his stay at UFRGS, and the program ``Science
without Borders'' from the Minist\'erio da Ci\^encia, Tecnologia e
Inova\c{c}\~ao (MCTI) and Minist\'erio da Educa\c{c}\~ao (MEC), Brazil
for generous financial support.

\end{acknowledgements}

\end{document}